\documentclass[12pt]{iopart}

\usepackage{iopams}

\usepackage{amsfonts}
\usepackage{setstack}
\usepackage{url}
\usepackage{amsthm}
\usepackage{graphicx}

\newtheorem{teo}{Theorem} \newtheorem{lemma}[teo]{Lemma}
\newtheorem{prop}[teo]{Proposition}

\def\endproof{\vrule height6pt width6pt depth0pt}

\begin{document}


\title[Bell inequalities from variable elimination methods]{Bell inequalities from variable elimination methods}


\author{Costantino Budroni}
\address{Departamento de F\'{\i}sica Aplicada II, Universidad de
 Sevilla, E-41012 Sevilla, Spain}
\ead{cbudroni@us.es}
\author{Ad\'an Cabello}
\address{Departamento de F\'{\i}sica Aplicada II, Universidad de
 Sevilla, E-41012 Sevilla, Spain}
\address{Department of Physics, Stockholm University, S-10691
 Stockholm, Sweden}
\ead{adan@us.es}


\begin{abstract}
Tight Bell inequalities are facets of Pitowsky's correlation polytope and are usually obtained from its extreme points by solving the hull problem. Here we present an alternative method based on a combination of algebraic results on extensions of measures and variable elimination methods, e.g., the Fourier-Motzkin method. Our method is shown to overcome some of the computational difficulties associated with the hull problem in some non-trivial cases. Moreover, it provides an explanation for the arising of only a finite number of families of Bell inequalities in measurement scenarios where one experimenter can choose between an arbitrary number of different measurements.
\end{abstract}





\section{Introduction}


Bell inequalities \cite{Bell64} play a fundamental role in the investigation of non-classical features of quantum mechanics (QM), from the foundational problems raised by the impossibility of a local hidden variable description of quantum systems, to applications for quantum information processing like entanglement-based reduction of classical communication complexity \cite{BZPZ04,BCMD10}, device-independent secure communication \cite{BHK05,ABGMPS07} and randomness expansion \cite{PAMBMMOHLMM10}. Similar inequalities, called non-contextuality inequalities, arise in the discussion of quantum contextuality \cite{KCBS08,Cabello08,BBCP09}.

In fact, the origin of all these inequalities is rooted in logic and probability \cite{Fine82,Pitowsky86,Pitowsky89} and can be traced back \cite{Pitowsky94} to Boole's notion of conditions of possible experience \cite{Boole62}; they are necessary conditions for the interpretation of a set of experimentally observed relative frequencies as probabilities in a single probability space.

From an algebraic point of view \cite{BM10}, Bell and non-contextuality inequalities can be seen as conditions for the extension of a function, defined on a subset of a Boolean algebra, to a normalized measure on the entire Boolean algebra, a problem already investigated by Horn and Tarski \cite{HT48}.

A general method for the derivation of Bell and non-contextuality inequalities is based on Pitowsky's notion of correlation polytope \cite{Pitowsky86}. Such a polytope appears in many fields and with different names (see \cite{DL97} for a extensive survey of the topic). Its geometrical structure is completely defined by a set together with a family of its subsets.

In the framework of QM, the above set is a set of observables and the subsets correspond to subsets of compatible or jointly measurable observables. The operational definition of such notions is still under discussion \cite{GKCLKZGR10}. They have been defined in different ways according to the experimental scenario and the physical assumptions involved (e.g., space-like separated measurements, sequential measurements). Such distinctions are irrelevant for our discussion. On the contrary, the use of an abstract notion of compatibility allows us for a unified presentation of our mathematical results which are valid in all the above mentioned scenarios.

Given a set of observables together with the subsets of compatible observables, the corresponding correlation polytope is defined as the convex hull of a set of vectors, the vertices or extreme points of the polytope, representing possible truth assignments for observables and logical conjunctions between compatible pairs, triples, etc.

Every convex polytope has two representations: One as the convex hull of its vertices ($\mathcal{V}$-representation) and the other as the intersection of a finite number of half-spaces ($\mathcal{H}$-representation), each given by a linear inequality. Complete sets of tight (i.e., facet supporting) Bell inequalities associated with a given set of observables precisely amount to the $\mathcal{H}$-representations of the corresponding correlation polytopes.

The $\mathcal{H}$-representation of a convex polytope can be computed starting from its $\mathcal{V}$-representation by solving the hull problem. For high-dimensional polytopes, this is a very difficult problem. In fact, this method has been used only to compute simple cases \cite{PS01,CG04}, by means of programs such as \texttt{cdd} \cite{Fukuda}, \texttt{lrs} \cite{lrs} and \texttt{porta} \cite{Porta}.

More recently, an alternative approach has been proposed by Avis, Imai, Ito and Sasaki \cite{AIIS04,AIIS05}, which is based on the relation between correlation polytopes and a family of convex polytopes called cut polytopes, well studied objects in polyhedral combinatorics, and a variable elimination method derived from Fourier-Motzkin method of elimination of variables (see, e.g., \cite{Ziegler95}) and called triangular elimination. Their method provided more than two hundred millions new tight Bell inequalities for cases in which the corresponding hull problem is computationally intractable. However, since such a method is based on cut polytopes, it can be applied only to the bipartite case.

Here we present an alternative general method based on the application of Fourier-Motzkin variable elimination method to conditions derived in Ref. \cite{BM10} as consistency conditions for putting together partial extensions of quantum probabilities in order to obtain a classical probability description.
More precisely, such conditions are expressed in terms of a systems of linear inequalities where also correlations between incompatible observables appear as variables: A classical probability space representation exists for a given set of QM predictions if and only if the corresponding system of linear inequalities admits a solution; Bell, or non-contextuality, inequalities are obtained by eliminating the variables associated to correlations between incompatible observables. Our approach can be seen as a generalization of Fine's derivation of the Clauser-Horne-Shimony-Holt (CHSH) polytope \cite{Fine82}.

The above method also provides a generalization of the result obtained by Sliwa \cite{Sliwa03} and Collins and Gisin \cite{CG04}, namely, the appearance of only a finite number of families of Bell inequalities in measurement scenarios where one experimenter is allowed to choose between an arbitrary number of different measurements.

The variable elimination method can also be applied to a family of correlation polytopes called complete probability polytopes. In this case, the method may not provide a significant computational advantage with respect to other known algorithms. Nevertheless, we believe that such results are interesting and we collected them in \ref{app:A}.

The paper is organized as follows: In Sec. \ref{sec:1}, the notion of correlation polytope is reviewed, together with the result on extensions of probabilities associated with tree graphs. In Sec. \ref{sec:2}, we discuss their application to the problem of the computation of $\mathcal{H}$-representation for general correlation polytopes, and provide some examples. Finally, in Sec. \ref{sec:3} we discuss some possible further developments and computational results.


\section{Correlation polytopes and extensions of measures}
 \label{sec:1}


In this section, we recall the definition of correlation polytope and its probabilistic interpretation; then we introduce the notion of compatibility graph for a set of observables and present the main result about the classical probability space representation for the corresponding QM predictions.

We use the definition of correlation polytope given in \cite{DL97}, which is the natural generalization of Pitowsky's notion \cite{Pitowsky89} to higher order correlations, but with a slightly different notation.

Given a set of propositions $\mathcal{G}=(A_1,\ldots,A_n)$ and a family $\mathcal{I}$ of subsets of $\mathcal{G}$, i.e., $\mathcal{I}\subset 2^{\mathcal{G}}$, we define the sets $S_k$, $k=2,\ldots,m$ with $m\leq n$, as the sets of logical conjunctions ${S_k=\{ A_{i_1}\wedge\cdots\wedge A_{i_k} \ |\ i_j\neq i_{j'},\ \{ A_{i_1},\ldots,A_{i_k}\}\in\mathcal{I}\ \}}$. The lines of the truth table associated to the above set of propositions and logical conjunctions between them, namely the $2^n$ vectors of $\mathbb{R}^{|\mathcal{G}|+|S_2|+\ldots+|S_m|}$,
\begin{equation}
 u_\varepsilon = (\varepsilon_1,\ldots,\varepsilon_n,\ldots,\varepsilon_i\varepsilon_j,\ldots, \varepsilon_{i_1}\varepsilon_{i_2}\cdots\varepsilon_{i_m},\ldots),
\end{equation}
where $\varepsilon=(\varepsilon_1,\ldots,\varepsilon_n)\in\{0,1\}^n$, are called the vertices of the correlation polytope. Their convex hull, i.e., the set of points generated by their convex combinations, is called the correlation polytope associated to $\mathcal{I}$ and denoted as $COR^{\square}(\mathcal{I})$.

It is convenient to introduce the following notation which makes apparent the correspondence between coordinates and joint probabilities. The coordinates of a point $p\in\mathbb{R}^{|\mathcal{G}|+|S_2|+\ldots+|S_m|}$ will be denoted as
\begin{equation}
 p=(p_1,\ldots, p_n, \ldots p_{ij}, \ldots, p_{i_1\ldots i_m},\ldots).
\end{equation}

In order to clarify our notation, it is convenient to present a simple example, the CHSH scenario. Consider the bipartite scenario in which one experimenter, say Alice, can choose between two dichotomic measurements, associated with propositions $A_1$ and $A_2$, and another experimenter, say Bob, can choose between two dichotomic measurements, associated with propositions $A_3$ and $A_4$. The set of propositions is, therefore, $\mathcal{G}=(A_1, A_2, A_3, A_4)$. Moreover, the measurements associated with $A_i$ and $A_j$, for $i=1,2$ and $j=3,4$, can be performed jointly and, consequently, it makes sense to consider the following set of logical conjunctions $S_2=\{ A_1 \wedge A_3, A_1 \wedge A_4, A_2 \wedge A_3, A_2 \wedge A_4 \}$. The associated polytope is described by $2^4=16$ vertices in $\mathbb{R}^8$, namely
\begin{equation}\label{eq:vch}
u_\varepsilon = (\varepsilon_1,\ \varepsilon_2,\ \varepsilon_3,\ \varepsilon_4,\ \varepsilon_1\varepsilon_3,\ \varepsilon_1\varepsilon_4,\ \varepsilon_2\varepsilon_3,\ \varepsilon_2\varepsilon_4),\qquad \varepsilon_i\in\{0,1\},
\end{equation}
where $\varepsilon_i$ represents a classical $\{0,1\}$-valued assignment to proposition $A_i$ and $\varepsilon_i \varepsilon_j$ the classical assignment for the logical conjunction $A_i\wedge A_j$.

As a consequence of Weyl-Minkowski theorem (see, e.g., \cite{Pitowsky89}), each convex polytope has a double description: One as the convex hull of its vertices $u_\varepsilon$, i.e., the $\mathcal{V}$-representation, and one as a (finite) intersection of half-spaces which generates it, each one given by a linear inequality, i.e., the $\mathcal{H}$-representation.

The convex hull of vertices (\ref{eq:vch}) gives a set of linear inequalities constraining the coordinates of a generic point in $\mathbb{R}^8$, which we denote by
\begin{equation}
 p=(p_1, p_2, p_3, p_4, p_{13}, p_{14}, p_{23}, p_{24}).
\end{equation}
Such inequalities, known as CHSH inequalities, are presented in eqs. (\ref{eq:ch})--(\ref{eq:ch4}) below and will be discussed in more detail.
The interpretation of such a geometrical object is the following: Given a vector belonging to the polytope, each component represents the joint probability for the corresponding subset of propositions, e.g., $p_{13}$ represents the joint probability ${Prob(A_1 \wedge A_3)}$; the whole polytope gives possible ranges for such joint probabilities if the underlying probabilistic structure is assumed to be classical, i.e., given by a probability space or, equivalently (since the number of propositions is finite) by a normalized measure on a Boolean algebra.

It is interesting to define the complete probability polytope associated to $n$ propositions ${\mathcal{G}=\{A_1,\ldots,A_n\}}$, it is the correlation polytope ${COR^\square(2^\mathcal{G})}$, i.e., the correlation polytope for a set of compatible proposition, in particular, ${|\mathcal{G}|+\sum_k |S_k|=\sum_{k=1}^n {n \choose k} =2^n-1}$.

The importance of such polytopes has been already recognized in \cite{DL97}, namely, the fact that each $COR^\square(\mathcal{I})$ can be obtained as a projection of $COR^\square(2^\mathcal{G})$ (on the subspace $\mathbb{R}^{|\mathcal{I}|}$). A more detailed discussion, containing also an algebraic derivation of the $\mathcal{H}$-representation for complete probability polytopes, can be found in \ref{app:A}.

We now recall a result on extension of probability measures which allows for another projection-based approach to the computation of $\mathcal{H}$-representation of correlation polytope. More details can be found in Ref.\cite{BM10}.


\begin{prop}\label{prop:tree1} (Representation of tree graphs)
Consider a set of probabilities $p_i$ on a set of $\{0,1\}$-valued
observables $A_i$ and correlations $p_{ij}$ on a subset of pairs $A_i, A_j$,
defining a probability on each pair
$A_i, A_j$, with $ p_i = \langle A_i \rangle$,
$ p_j = \langle A_j \rangle$, $ p_{ij} = \langle A_i A_j \rangle$; now depict observables $A_i$ as vertices and the above pairs as edges in a graph. Then any set of predictions associated with \textit{a tree graph}, i.e., a graph without closed loops, admits a classical representation.
\end{prop}


\begin{prop}\label{prop:tree2} The same holds with
\textit{yes/no observables} $A_i$ substituted by
free Boolean algebras $ \mathcal{A}_i $,
$p_i$ by probabilities on
$ \mathcal{A}_i $, $p_{ij} $ by probabilities on the Boolean algebra freely
generated by the union of the sets of generators of
$ \mathcal{A}_i $ and $ \mathcal{A}_j $.
\end{prop}


We recall that a Boolean algebra is freely generated by $n$ generators $B_1,\ldots,B_n$ if such generators are as much unconstrained as possible, i.e., they satisfy no conditions except those necessary conditions defining a Boolean algebra (e.g., distributive law). Since all Boolean algebras freely generated by $n<\infty$ generators are isomorphic, for the sake of simplicity, we can think of the the algebra of subsets $2^{X}$ of the set $X=\{0,1\}^n$, with set theoretic operations $(\cap, \cup, ^c)$; then the subsets ${B_i=\{ (x_1,\ldots,x_n)\in X\ |\ x_i = 1\ \}}$, for $i=1,\ldots, n$, can be taken as free generators. In terms of propositions and truth assignments, the free Boolean algebra assumption amounts to the assumption that each possible $\{0,1\}$-valued assignment to propositions is admissible. For more details see \cite{GH09}. Notice that this is precisely the way how we define the $2^n$ vertices of the correlation polytope.

In the next section, we shall show how to exploit the above results to obtain $\mathcal{H}$-representation for general correlation polytopes starting from lower dimensional polytopes and using Fourier-Motzkin method.


\section{Derivation of Bell inequalities}
 \label{sec:2}


The above results provide a general method for the computation of $\mathcal{H}$-representation for correlation polytopes which consists in solving the hull problem for a smaller polytope, then constructing an higher dimensional polytope from that solution, and applying variable elimination methods such as, e.g., the Fourier-Motzkin method.

For the convenience of the reader we recall briefly the Fourier-Motzkin method, for more details see \cite{Ziegler95}. Consider a system of inequalities of the form $Ax\leq b$, where $A$ is a $m\times d$ real matrix, $x=(x_1,\ldots,x_d)\in \mathbb{R}^d$ and $b\in\mathbb{R}^m$, and suppose we want to eliminate the variable $x_d$. After a proper normalization, each inequality will be of the form $a_{i1}x_1+\ldots+a_{id} x_d\leq b_i$ with $a_{id}$ being $+1,-1$ or $0$. By summing
an inequality where $a_{id}=+1$ with an inequality where $a_{jd}=-1$, we obtain a new inequality not containing $x_d$. By repeating the above operation for every possible pair $i,j$ such that $a_{id}=+1$ and $a_{jd}=-1$, and considering also the inequalities where
$a_{id}=0$, we obtain a new system of inequalities not containing $x_d$.
From a geometric point of view, since the system of linear inequalities can represent a polytope (more generally, a cone), the above operation amounts to a projection on the coordinates associated with the variables $x_1,\ldots,x_{d-1}$.

Our method consists in exploiting the automatically existing classical representations for subsets of observables with compatibility relations described by tree graphs, see Propositions \ref{prop:tree1} and \ref{prop:tree2} above. Conditions for classical representability arise as consistency (i.e., coincidence on intersections) conditions for putting together partial extensions associated with subgraphs, giving rise to a description of the initial compatibility graph as a tree graph on such extended nodes. Such consistency conditions are expressed in terms of the existence of a solution for a set of linear inequalities.
One of the main application of Fourier-Motzkin algorithm is precisely deciding whether a system of inequalities has a solution.

Moreover, we shall discuss how this result gives an account for the appearance of a finite number of families of Bell inequalities in cases in which one experimenter can choose between arbitrary number of different measurements while such a number is fixed for the others, as recognized by Sliwa \cite{Sliwa03} and Collins and Gisin \cite{CG04} for the case in which Alice perform two measurement and Bob an arbitrary number.

We recall that, although our method is general, different strategies are possible corresponding to different partitions of the initial graph into subgraphs. We shall discuss our method by means of some simple examples.
The first one is the derivation of the CHSH polytope. It is interesting to notice that it is analogous to that presented by Fine \cite{Fine82}; our method can be seen as a generalization of his idea to an arbitrary number of observables.


\subsection{CHSH polytope from Bell-Wigner polytope}\label{sec:2.1}


\begin{figure}[t]
\begin{center}
\includegraphics[width=13.6cm]{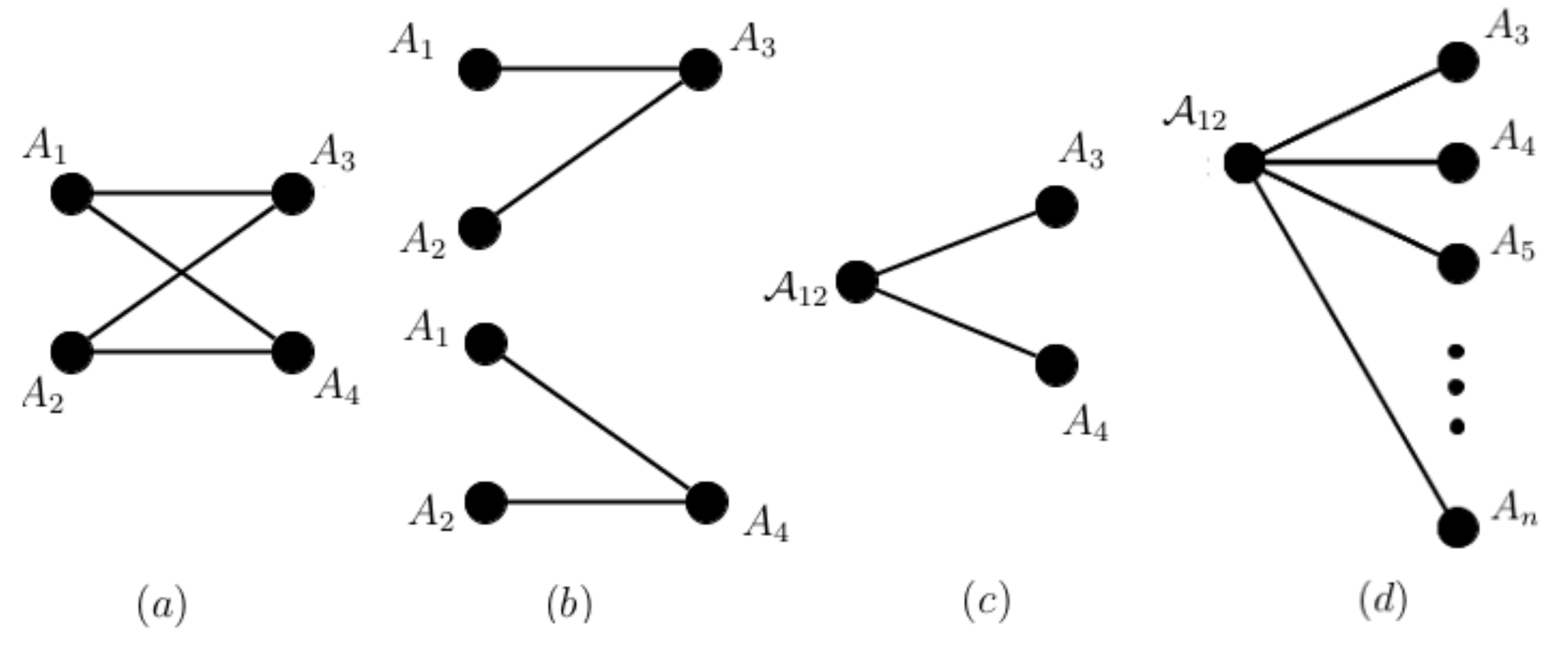}
\caption{\label{Fig1} (a) Graph of the compatibility relations between the observables in the CHSH scenario. (b) Partition in two subset of three observables with intersection $\{A_1, A_2\}$. (c) Tree graph obtained by extending the probability measure on the algebra generated by $A_1, A_2$. (d) Asymmetric case with additional observables on Bob's side.}
\end{center}
\end{figure}


The CHSH polytope is generated by the following set of vertices
\begin{equation}
u_\varepsilon = (\varepsilon_1,\ \varepsilon_2,\ \varepsilon_3,\ \varepsilon_4,\ \varepsilon_1\varepsilon_3,\ \varepsilon_1\varepsilon_4,\ \varepsilon_2\varepsilon_3,\ \varepsilon_2\varepsilon_4),\qquad \varepsilon_i\in\{0,1\}.
\end{equation}
It is associated with a bipartite measurement scenario in which Alice can choose between two measurements, associated with propositions $A_1$ and $A_2$, and Bob can choose between two measurements, associated with propositions $A_3$ and $A_4$.

As already recognized by Fine \cite{Fine82} and discussed also in \cite{BM10,BM11}, the existence of a classical description for the four observables is equivalent to the existence of classical descriptions for the two subsystems, $\{A_1, A_2, A_3\}$ and $\{A_1, A_2, A_4\}$, coinciding on $\{A_1, A_2\}$. In fact, the two classical descriptions would give rise to an extension of the probability assignment to the four observables satisfying the hypothesis of Proposition \ref{prop:tree2} (see Figure \ref{Fig1} (a),(b),(c)).

The constraints on the subsystem $\{A_1, A_2\}$ imposed by the third observable, $A_3$ or $A_4$, are described by the Bell-Wigner polytope, i.e., the correlation polytope associated to three propositions and their pairwise logical conjunctions, which is given by the following inequalities, obtained in Ref.\cite{Pitowsky89},
\numparts
\begin{eqnarray}
 \label{eq:bw1}
0\leq p_{ij}\leq p_i,\;\; 0\leq p_{ij}\leq p_j,\;\; ij=12, 1s, 2s,\\
\label{eq:bw2}
p_i+p_j-p_{ij}\leq 1,\;\; ij=12, 1s, 2s,\\
\label{eq:bw3}
1-p_1-p_2-p_s+p_{12}+p_{1s}+p_{2s}\geq 0,\\
\label{eq:bw4}
p_1-p_{12}-p_{1s}+p_{2s}\geq 0,\\
\label{eq:bw5}
p_2-p_{12}-p_{2s}+p_{1s}\geq 0,\\
\label{eq:bw6}
p_s-p_{1s}-p_{2s}+p_{12}\geq 0,
\end{eqnarray}
\endnumparts
for $s=3$ or $4$. For fixed $s$ $(3$ or $4)$, the above inequalities give ranges for the classical probabilities for a system of three proposition and their pairwise joint probabilities. It is interesting to notice that, by Proposition \ref{prop:tree1}, each subset of three observables always admit a classical representation (see also Figure \ref{Fig1} (b)); therefore, for fixed $s$, the above system of inequalities admits a solution, i.e., a value can be assigned to the joint probability $p_{12}$ (which is assumed to be experimentally unmeasurable, since it is associated with two noncommuting observables) consistently with the constraints imposed by the classical model and the measurable joint probabilities $p_{1s}$ and $p_{2s}$. This implies that, for fixed $s$, the elimination of variable $p_{12}$ gives rise only to trivial inequalities, i.e., inequalities always satisfied by quantum predictions. An analysis of the constraints imposed on such ``experimentally unmeasurable'' joint probabilities has been presented in \cite{BM11}.
Analogous situations always arise in the bipartite scenario and this additional property just gives more information about the triviality of certain inequalities. However our method can also be applied to the general multipartite case.

Classical representability for QM prediction in the CHSH scenario amounts, therefore, to the existence of a common solution for the two systems of linear inequalities, namely, it amounts to the existence of a value for $p_{12}$ consistent with the constraints imposed by classical descriptions for the two subsystems of three observables.

As a consequence of general properties of Fourier-Motzkin method (see \cite{Ziegler95}), a system of inequalities admits a solution if and only if the projected system, i.e., the system obtained by eliminating one or more variables, admits a solution. It follows that the above set of measurements admits a classical description if and only if measured correlations, i.e., correlations between compatible observables, satisfy the system of inequalities obtained by eliminating $p_{12}$.

In order to eliminate $p_{12}$, we just combine inequalities where $p_{12}$ appears with opposite sign and keep inequalities where it does not appear. As discussed before, by Proposition \ref{prop:tree1}, combining inequalities with the same index $s$ gives rise only to trivial inequalities, in particular, they are also redundant. A posteriori, we know that the same happens also for inequalities (\ref{eq:bw1}) and (\ref{eq:bw2}).

The interesting inequalities are those obtained from the combination of (\ref{eq:bw3})--(\ref{eq:bw6}) for different $s$, namely
\numparts
\begin{eqnarray}
\label{eq:ch}
-1\leq p_{13}+p_{14}+p_{24}-p_{23}-p_1-p_4\leq 0,\\
-1\leq p_{23}+p_{24}+p_{14}-p_{13}-p_2-p_4\leq 0,\\
-1\leq p_{14}+p_{13}+p_{23}-p_{24}-p_1-p_3\leq 0,\\
\label{eq:ch4}
-1\leq p_{24}+p_{23}+p_{13}-p_{14}-p_2-p_3\leq 0.
\end{eqnarray}
\endnumparts
For instance, $-1\leq p_{13}+p_{14}+p_{24}-p_{23}-p_1-p_4$ is obtained as a sum of inequality (\ref{eq:bw3}) for $s=4$, which contains the term $+p_{12}$, with inequality (\ref{eq:bw5}) for $s=3$, which contains the term $-p_{12}$.

Inequalities (\ref{eq:ch})--(\ref{eq:ch4}), together with the inequalities (\ref{eq:bw1}), (\ref{eq:bw2}) in which $p_{12}$ does not appear, give the $\mathcal{H}$-representation of the CHSH polytope (compare with, e.g., \cite{Pitowsky89}).


\subsection{Bipartite $(2,n)$ scenario}


An analogous argument applies to the scenario in which Alice can choose between two measurements and Bob can choose among $n>2$ measurements, associated to propositions $A_3,\ldots,A_{n+2}$: The initial system of inequalities is still given by (\ref{eq:bw1})--(\ref{eq:bw6}), but with $s$ taking values in $\{3,4,\ldots,n+2\}$ (see Figure \ref{Fig1} (d)).

Since only one variable (i.e., $p_{12}$) has to be eliminated, at most two inequalities with different index $s$ can be combined to give a valid inequality. Therefore, the final set of inequalities is given by (\ref{eq:ch})--(\ref{eq:ch4}) with the pair $3,4$ substituted by any pair $i,j$ with $i,j\in\{3,\ldots,n+2\}$ and $i<j$. This is precisely the result obtained in Refs. \cite{CG04,Sliwa03}.


\subsection{Two parties, three settings}
\label{subs33}


Now consider the bipartite scenario in which Alice can choose among three measurements, associated with propositions $A_1, A_2$ and $A_3$, and Bob can choose among three measurements, associated with propositions $A_4, A_5$ and $A_6$.

\begin{figure}[t]
\begin{center}
\includegraphics[width=11.0cm]{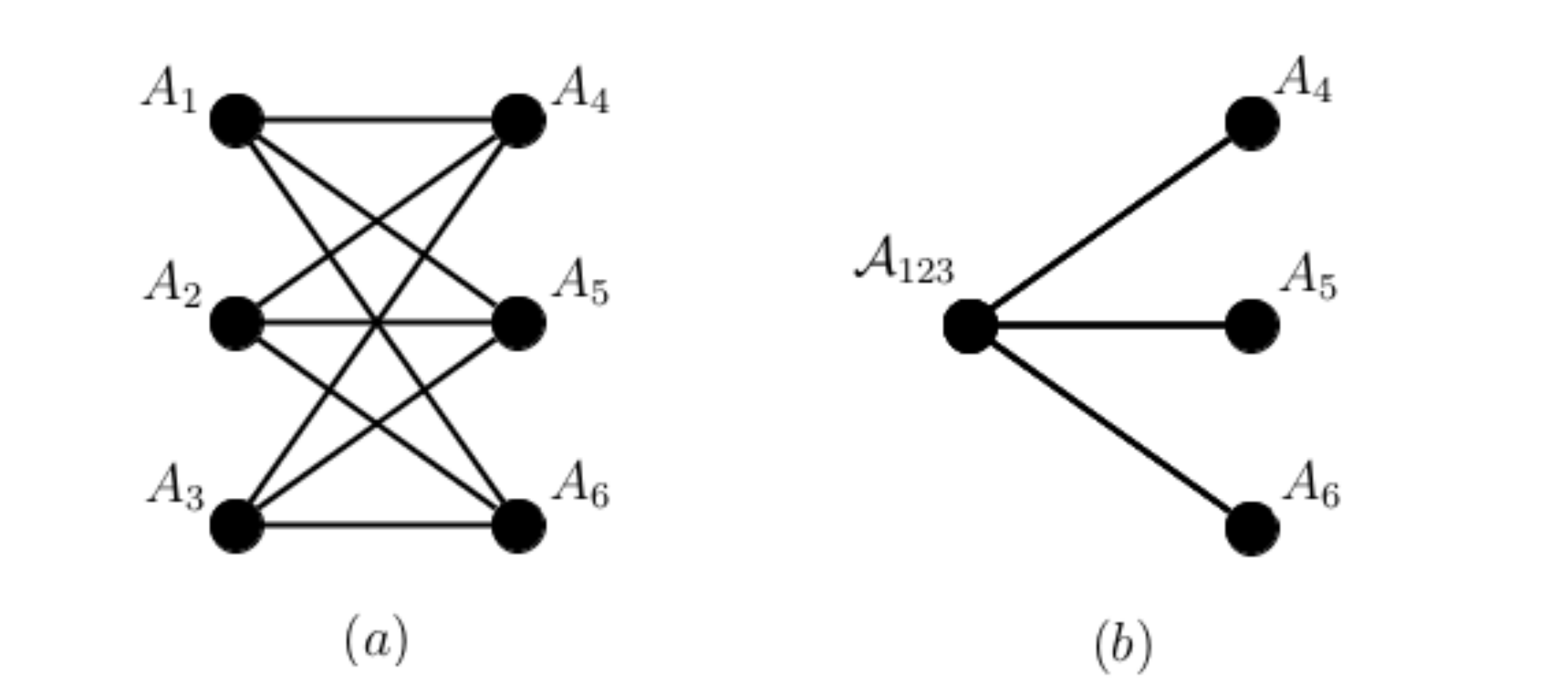}
\end{center}
\caption{\label{Fig2} (a) Graph of the compatibility relations between the observables in the bipartite (3,3) scenario. (b) Tree graph obtained by extending the measure on the algebra generated by $A_1, A_2, A_3$. }
\end{figure}

Analogously to the previous case, see Figure \ref{Fig2}, the existence of a classical description for the six observables is equivalent to the existence of classical descriptions for the three subsystems, $\{A_1, A_2, A_3, A_4\}$, $\{A_1, A_2, A_3, A_5\}$ and $\{A_1, A_2, A_3, A_6\}$, coinciding on $\{A_1, A_2, A_3\}$. A probability on $\{A_1, A_2, A_3\}$ is completely defined, see Lemma \ref{lemma:misunic} below, once the probabilities $p_1, p_2, p_3, p_{12}, p_{13}, p_{23}, p_{123}$ are given.

It is therefore sufficient to calculate the correlation polytope associated with probabilities $p_1, p_2, p_3, p_s, p_{1s}, p_{2s}, p_{3s}, p_{12}, p_{13}, p_{23}, p_{123}$, then consider the system given by all the above inequalities for $s=4,5,6$ and eliminate the variables
$p_{12}, p_{13}, p_{23}, p_{123}$.


\subsection{Bipartite $(3,n)$ scenario}


Again, by adding observables only on Bob's side, one just obtains more copies of the initial system of inequalities, but with different indices $s$. The situation is analogous to that depicted in Figure \ref{Fig1} (d), but with $\mathcal{A}_{12}$ substituted by $\mathcal{A}_{123}$

In particular, since four variables have to be eliminated, at most $2^4=16$ inequalities with different indices $s$ can be combined. As a result, all families of valid inequalities for the general case $(3,n)$ already arise in the case in which Alice performs 3 measurement and Bob 16.


\subsection{Multipartite $(m,\ldots,m,n)$ scenario}


The above argument can be extended to the case of $p$ parties in which the first $p-1$ can choose among $m$, with $m$ fixed, measurements, while the last one can choose among $n>m$, with $n$ arbitrary, measurements: All families of inequalities can be obtained by studying the case in which the last experimenter performs $2^k$ measurements, where $k$ is the number of variables to be eliminated.


\section{Conclusions and computational results}
 \label{sec:3}


We have presented an alternative method for the computation of half-space representation for correlation polytopes based on algebraic conditions and variable elimination. A reasonable question is: Does it provide any advantage with respect to existing methods? In order to show the advantages of the tree graph method, we have computed the $\mathcal{H}$-representation for some simple polytopes, with (i) our tree graph method using existing software implementing the Fourier-Motzkin algorithm; specifically, we used \texttt{porta} \cite{Porta} and \texttt{FM} library \cite{FMlib}, and (ii) using standard software for solving the hull problem; specifically, we used \texttt{cdd}.

For simple cases like the $(2,2)$ (i.e., the CHSH) and $(3,3)$ scenarios, the computation is equally fast with both methods. However, remarkably, our tree graph method is noticeably faster to compute asymmetric scenarios:

For the $(3,4)$ scenario, the tree graph method implemented with \texttt{porta} completed the calculation in $\approx$ 11 minutes, while \texttt{cdd} needed $\approx$ 20 minutes. The 11 minutes include the time (seconds) required to calculate the initial polytope (see Sec. \ref{subs33}).

For the $(3,5)$ scenario, the tree graph method implemented with \texttt{porta} completed the calculation in $\approx$ 72 minutes, while \texttt{cdd} was still running after a week and we had to stop it.
All computations were performed on the same machine with an Intel Xeon CPU running at $3.20$ GHz.

However, beyond the practical advantage the tree graph method has to compute some scenarios, the main purpose of this paper is to point out that, for some polytopes, the $\mathcal{H}$-representation can be easily expressed by means of simple algebraic arguments, and then these polytopes can be related to those associated to polytopes arising in physically interesting scenarios by means of variable elimination methods. For example, the Boolean algebraic approach in Ref. \cite{HT48} provided, already in 1948, complete sets of Bell inequalities. To our knowledge, this approach has never been investigated in relation to Bell inequalities, and we hope that our work will stimulate further developments in this direction.


\ack
The authors thank Louis-N\"{o}el Pouchet for helpful discussions about his software and Fourier-Motzkin algorithm in general, Emilio Pinna for his help in writing some of the programs used for the calculations, and an anonymous referee for useful comments and suggestions. This work was supported by the Spanish Projects FIS2008-05596 and FIS2011-29400, and the Wenner-Gren Foundation.


\appendix


\section{}\label{app:A}


In this appendix we recall some results, Proposition \ref{prop:polit} and Lemmas \ref{lemma:normmeas}, \ref{lemma:misunic}, arising from the translation of Pitowsky's notion of correlation polytope into the Boolean framework, and use them to derive the $\mathcal{H}$-representation for the complete correlation polytope. For more details on such results and the basic notions of Boolean algebra involved, see \cite{BM10}.

After completing our work, we realized that $i)$ obtaining a correlation polytope as a projection of the corresponding complete probability polytope need approximatively the same amount of time and resources as the usual convex hull algorithms based on Fourier-Motzkin elimination, and, $ii)$ similar results were already known in the field of integer $0/1$ programming. In fact, in Ref.\cite{LR05}, the $\mathcal{H}$-representation of the complete probability polytope is obtained by finding the explicit form of the matrix $N$ we defined in Lemma \ref{lemma:full} below. Nevertheless, we believe our approach is interesting, since it is based on logical and probabilistic, rather than geometrical, notions and provides an alternative point of view which has not been explored, to our knowledge, in the context of Bell inequalities.

First we present the $\mathcal{H}$-representation for the complete probability polytope associated with $n$ propositions. It is given by the following set of $2^n$ tight inequalities:
\begin{equation}
 \label{eq:he}
 h(\varepsilon_1,\ldots,\varepsilon_n)\geq 0, \quad \varepsilon=(\varepsilon_1,\ldots,\varepsilon_n)\in\{0,1\}^n,
\end{equation}
where $h(\varepsilon_1,\ldots,\varepsilon_n)$ is defined as
\begin{equation}
 \label{eq:expmis0}
\eqalign{
 h(\overbrace{1,\ldots,1}^{s},\overbrace{0,\ldots, 0}^{k})=p_{1\cdots s} -\sum_{_{s+1\leq l\leq s+k}} p_{1\cdots sl}
 +\sum_{_{s+1\leq l_1<l_2\leq s+k}} p_{1\cdots sl_1l_2}+\cdots\\
 \cdots + (-1)^{k-1} \sum_{_{s+1\leq l_1<l_2<\cdots<l_{k-1}\leq s+k}} p_{1\cdots sl_1l_2\cdots l_{k-1}}+
 (-1)^k p_{1\cdots n},}
\end{equation}
for all $1\leq s \leq n$ and $s+k=n$; all other values, except $h(0,\ldots,0)$, are defined by a permutation of indices $\{1,\ldots,n\}$ providing the above form, i.e., $(1,\ldots,1,0,\ldots,0)$, for $(\varepsilon_1,\ldots,\varepsilon_n)$; in the case $s=0$, the first term on the r.h.s is $1$.

The notion of correlation polytope has been introduced by Pitowsky \cite{Pitowsky89} in terms of propositional logic, however the same problem can be expressed in terms of free Boolean algebras with the identification (see \cite{GH09}) of atomic proposition with free generators, logic operations with Boolean operations, truth assignments with two-valued measures and probability assignments with normalized measures.

To prove that the above representation is correct, we first need the following basic fact


\begin{lemma}\label{lemma:normmeas}
 Let $\mathfrak{B}$ be a finite Boolean algebra with $k$ atoms
 $\{a_1,\ldots,a_k\}$; then a function $f:X\subset\mathfrak{B}\longrightarrow [0,1]$ can be extended to normalized measure $\mu$ on
$\mathfrak{B}$ if and only if there are $k$ numbers
$\lambda_1,\ldots,\lambda_k$, with $\lambda_i\geq 0$ and $\sum_{i=0}^k
\lambda_i =1$, such that
\begin{equation}
 \label{eq:est}
 f=\sum_{i=1}^k \lambda_i {\delta_{a_i}}_{|_X},
 \end{equation}
where $\delta_{a_i}$ is the two-valued measure which is
 $1$ on $a_i$. $\mu$ is given by
\begin{equation}
 \mu=\sum_{i=1}^k \lambda_i \delta_{a_i},
\end{equation}
and the coefficients correspond to measure of atoms, i.e., $\lambda_i=\mu(a_i)$.
\end{lemma}


As a consequence, Pitowsky's result can be expressed as follows:


\begin{prop}\label{prop:polit}
Let $\mathfrak{B}$ be a Boolean algebra freely generated by
$\mathcal{G}=\{A_1,\ldots,A_n\}$ and consider $X\subset \mathfrak{B}$ with
$X=\{A_1,\ldots,A_n,\ldots,A_i\cap A_j,\ldots,A_i\cap A_j\cap A_k,\ldots\}$,
i.e., $X=\mathcal{G}\cup S_2\cup\ldots S_m$, $m\leq n$, where elements of $S_l$
are the intersections of $l$ distinct generators, but not necessarily all of
those possible, i.e., $|S_l|\leq {n \choose l}$. Now consider
$f:X\longrightarrow[0,1]$ and define the vector
\begin{equation}
 p=(p_1,\ldots,p_n,\ldots p_{ij},\ldots,p_{i_1\ldots i_m},\ldots)\in
\mathbb{R}^{|X|},
\end{equation}
 which has as components the values assumed by $f$ on $X$, namely
\begin{eqnarray}
 p_i&=f(A_i),\quad i=1,\ldots,n,\\
 p_{ij}&=f(A_i\cap A_j),\quad A_i\cap A_j\in S_2,\ldots\\
p_{i_1\ldots i_m}&=f(A_{i_1}\cap\ldots \cap A_{i_m}),\quad A_{i_1}\cap\ldots\cap A_{i_m}\in S_m.
\end{eqnarray}
For every $\varepsilon \in \{0,1\}^n$ define the vector $u_{\varepsilon}\in
 \{0,1\}^{|X|}$ given by
\begin{equation}
 u_\varepsilon = (\varepsilon_1,\ldots,\varepsilon_n,\ldots,\varepsilon_i
 \varepsilon_j,\ldots,
 \varepsilon_{i_1}\varepsilon_{i_2}\ldots\varepsilon_{i_m},\ldots),
\end{equation}
i.e., for every component $p_{i_1\ldots i_k}$ of $p$ there is a corresponding component of $u_\varepsilon$ given by $\varepsilon_{i_1}\ldots\varepsilon_{i_k}$.

Then $f$ can be extended to a normalized measure on $\mathfrak{B}$ if and only if there are $2^n$ numbers $\lambda(\varepsilon)$, $\varepsilon\in\{0,1\}^n$,
such that
\begin{equation}
 p=\sum_{\varepsilon\in\{0,1\}^n} \lambda(\varepsilon) u_\varepsilon,
\end{equation}
with
\begin{equation}
 \lambda(\varepsilon) \geq 0 \text{ } \forall \varepsilon \in
 \{0,1\}^n,\text{ and }\sum_{\varepsilon \in \{0,1\}^n}
 \lambda(\varepsilon)=1 \ .
\end{equation}
\end{prop}


The convex hull of the vectors $u_\varepsilon$ defined above is known as the correlation polytope; clearly, each $A_i$ is associated to a QM observable, each set $S_k$ to a joint measurements of $k$ observables, and numbers $p_{i_1,\ldots,i_k}$ to probabilities for the corresponding measurements

If $X$ contains all possible intersections between generators, i.e., ${|X|=\sum_{k=1}^n {n \choose k} =2^n-1}$, it will be called a complete set of measurements and the corresponding polytope will be called complete probability polytope, and it will be denoted as $\mathcal{CP}_n$; such names come from the fact that there is a bijection between the points of the polytope and the normalized (i.e., probability) measures on the Boolean algebra freely generated by $A_1,\ldots,A_n$, as a consequence of the following general fact \cite{BM10}:


\begin{lemma}\label{lemma:misunic}
Let $\mathfrak{B}$ be a Boolean algebra freely generated by $\{A_1,\ldots,A_n\}$, let $\mu$ be a normalized measure on $\mathfrak{B}$ and let $X\subset \mathfrak{B}$ be a complete set of measurements. Then the measure $\mu$ is uniquely defined by the values it assumes on the set $X$.
\end{lemma}

Now consider a Boolean algebra $\mathfrak{B}$ freely generated by $\{A_1,\ldots,A_n\}$, a complete set of measurement $X\subset \mathfrak{B}$ and a function $f:X\longrightarrow [0,1]$. If $f$ admits extension to a normalized measure $\mu$, then the measure of the atoms of $\mathfrak{B}$, i.e., the elements $A_1^{\varepsilon_1}\cap A_2^{\varepsilon_2}\cap\ldots\cap A_n^{\varepsilon_n}$, where $\varepsilon_i\in\{0,1\}$ and $A_i^{\varepsilon_i}$ is defined as $A_i$ if $\varepsilon_i=1$ and $A_i^c$ (the Boolean complement) otherwise, can be written as
\begin{equation}\label{eq:misat}
\left. \begin{array}{l}
\mu(A_1\cap \ldots\cap A_n)=f(A_1\cap\ldots\cap A_n)\ ,\\
\mu(A_1\cap \ldots\cap A_n^c)=f(A_1\cap\ldots\cap A_{n-1})-f(A_1\cap\ldots\cap A_n)\ ,\\
\qquad \qquad \vdots \qquad\qquad \vdots\qquad \qquad \vdots \\
\mu(A_1^c \cap \ldots\cap A_n^c)=1-\sum_i f(A_i)+\sum_{i<j}f(A_i \cap A_j)\\
 -\sum_{i<j<k}f(A_i\cap A_j\cap A_k)+\ldots+(-1)^n f(A_1\cap\ldots\cap A_n) \ ,
\end{array}\right.
\end{equation}
where the above expressions are obtained by using the identities $\mu (A)=\mu(A\cap B)+\mu(A\cap B^c)$ for all $A,B\in\mathfrak{B}$ (additivity of measure on disjoint elements) and $\mu(\mathbf{1})=1$, $\mathbf{1}$ being the identity in $\mathfrak{B}$ (normalization).

The exact expression for the values of the measure on atoms in terms of values on intersections is given by the following general result:


\begin{lemma}\label{lemma:expmis}
Given a Boolean algebra $\mathfrak{B}$ and $n\geq 2$ elements $B_1,\ldots,B_n\in\mathfrak{B}$ and a measure $\mu$ on $\mathfrak{B}$, it holds
\begin{equation}\label{eq:expmis}
\eqalign{
\fl\mu(B_1\cap\ldots\cap B_s\cap B_{s+1}^c\cap\ldots\cap B_{s+k}^c)=\mu (B_1\cap\ldots\cap B_s)+\\
\fl-\sum_{_{s+1\leq l\leq s+k}} \mu(B_1\cap\ldots\cap B_s\cap B_l)
+\sum_{_{s+1\leq l_1<l_2\leq s+k}} \mu(B_1\cap\ldots\cap B_s\cap B_{l_1}\cap B_{l_2})
+\ldots\\
\fl\ldots + (-1)^{k-1} \sum_{_{s+1\leq l_1<l_2<\ldots<l_{k-1}\leq s+k}} \mu(B_1\cap\ldots\cap B_s\cap B_{l_1}\cap B_{l_2}\cap\ldots\cap B_{l_{k-1}})+\\
\fl +(-1)^k\mu(B_1 \cap \ldots\cap B_{s+k}) \ ,}
\end{equation}
with $s$ and $k$ such that $1\leq s\leq n$, and $k=n-s$. In the case $s=0$, i.e., $\mu(B_1^c\cap\ldots\cap B_n^c)$, an analogous expression holds, but with the first term $\mu(B_1\cap\ldots\cap B_s)$ substituted by $\mu(\mathbf{1})$, which is equal to $1$ if $\mu$ is a normalized measure.
\end{lemma}


\textbf{Proof} Fixed $n\geq 2$, we proceed by induction on $k$. For $k=0$ we have nothing to show. For $k=1$ it is sufficient to use the identity $\mu(A\cap B^c)=\mu (A)-\mu(A\cap B)$.

For the inductive step, it is sufficient to define $B'_1 = B_1 \cap B_{s+k}^c$, apply the inductive hypothesis to ${\mu(B'_1\cap B_2\cap\ldots\cap B_s\cap B_{s+1}^c\cap \ldots\cap B_{s+k-1}^c})$ and then apply again the identity $\mu(A\cap B^c)=\mu (A)-\mu(A\cap B)$ to eliminate $B_{s+k}^c$.

It is obvious that the above argument does not depend on $n$.
The proof for $s=0$ is analogous.\hfill \endproof


Now consider a generic function $f:X\rightarrow [0,1]$ and define $\widetilde{X}$ as the set of atoms $\widetilde{X}\equiv\{A_1^{\varepsilon_1}\cap A_2^{\varepsilon_2}\cap\ldots\cap A_n^{\varepsilon_n}\ |\ \varepsilon_i\in\{0,1\} \ \}$ and a function $\tilde{f}:\widetilde{X}\longrightarrow \mathbb{R}$ in terms of $f$ in analogy with the result of Lemma \ref{lemma:expmis}, i.e., for each $A_1^{\varepsilon_1}\cap A_2^{\varepsilon_2}\cap\ldots\cap A_n^{\varepsilon_n}$, $\tilde{f}$ is defined as the r.h.s. of the corresponding equation in (\ref{eq:misat}). Notice that by construction and independently of the values of $f$, in particular, independently of the existence of an extension for $f$, $\tilde{f}$ satisfies
\begin{eqnarray}
\label{eq:sum}
\fl \qquad \qquad f(A_{i_1}\cap \ldots \cap A_{i_k})&=\sum_{\varepsilon_{i_{k+1}},\ldots,\varepsilon_{i_n}}\tilde{f}(A_{i_1}\cap\ldots\cap A_{i_k}\cap A_{i_{k+1}}^{\varepsilon_{i_{k+1}}}\cap\ldots\cap A_{i_n}^{\varepsilon_{i_n}})\ ,\ \\
\label{eq:sum2} 1 &=\sum_{\varepsilon_{1},\ldots,\varepsilon_{n}}\tilde{f}(A_{1}^{\varepsilon_{1}}\cap\ldots\cap A_{n}^{\varepsilon_{n}})\ ,
\end{eqnarray}
where $(i_1,\ldots,i_n)$ is a permutation of $(1,\ldots,n)$. In particular, each of the two functions, $f$ and $\tilde{f}$, is completely determined by the other one.

Extensions of $f$ and $\tilde{f}$ are related by the following


\begin{lemma}\label{lemma:ff}
$f$ admits an extension to a measure $\Longleftrightarrow$ $\tilde{f}$ admits an extension to a measure. Moreover, if the two extensions exist, they coincide.
\end{lemma}
\textbf{Proof} The first implication follows from the definition of $\tilde{f}$; the other follows from Eq. (\ref{eq:sum}) together with the identity
\begin{equation}
A_{i_1}\cap \ldots \cap A_{i_k}=\bigcup_{\varepsilon_{i_{k+1}},\ldots,\varepsilon_{i_n}}A_{i_1}\cap\ldots\cap A_{i_k}\cap A_{i_{k+1}}^{\varepsilon_{i_{k+1}}}\cap\ldots\cap A_{i_n}^{\varepsilon_{i_n}} \ ,
\end{equation}
for all $(i_1,\ldots,i_n)$ permutation of $(1,\ldots,n)$ and all $1\leq k\leq n$, since the above is a disjoint union.

To conclude, it is sufficient to notice that a measure is completely defined by the values it assumes on the atoms of the Boolean algebra (since each element can be written in a unique way as as disjoint union of atoms) and that each of the two functions is uniquely determined by the other one.\hfill \endproof

We are now able to prove the main result


\begin{teo}\label{teo:hr} \textbf{($H$-representation of } $\mathcal{CP}_n$\textbf{)}\\
A vector $p=(p_1,\ldots,p_n,\ldots,p_{ij},\ldots,p_{i_1\ldots i_m},\ldots,p_{1\ldots n})\in
\mathbb{R}^{2^{n}-1}$ belongs to $\mathcal{CP}_n$ $\Longleftrightarrow$ its components satisfy the following $2^n$ inequalities
\begin{equation}\label{eq:hrep}
\left\lbrace\begin{array}{l}
0 \leq p_{1\ldots n} \\
0\leq p_{1\ldots (n-1)}-p_{1\ldots (n-1) n}\\
\qquad \qquad \vdots\\
0\leq 1-\sum_{i} p_i +\sum_{ij} p_{ij} -\sum_{ijk} p_{ijk} + \ldots
\end{array}\right.
\end{equation}
where the r.h.s. of inequalities (\ref{eq:hrep}) are obtained from the r.h.s of eqs. (\ref{eq:misat}) by substituting ${f(A_{i_1}\cap \ldots \cap A_{i_k})}$ with $p_{i_1\ldots i_k}$ for all $\{i_1,\ldots,i_k\}\subset\{1,\ldots,n\}$.
\end{teo}


\textbf{Proof} Given $p\in \mathbb{R}^{2^{n}-1}$, consider the function $f:X\subset \mathfrak{B}\longrightarrow \mathbb{R}$, where $\mathfrak{B}$ is the Boolean algebra freely generated by $\{A_1,\ldots,A_n\}$ and $X$ a complete set of measurements, defined as
\begin{equation}
f(A_{i_1}\cap\ldots\cap A_{i_k})=p_{i_1\ldots i_k}, \text{ for all } \{i_1,\ldots,i_k\}\subset\{1,\ldots,n\} \ .
\end{equation}
Let us assume that $f$ takes values in the interval $[0,1]$, we shall prove it later. Then, by Proposition \ref{prop:polit}, the above problem amounts to the problem of the existence of a measure extending $f$ on $\mathfrak{B}$, and therefore, by Lemma \ref{lemma:ff}, to the problem of the existence of a measure extending $\tilde{f}$. Since $\tilde{f}$ is defined on the atoms of $\mathfrak{B}$ and each element can be written uniquely as a disjoint union of atoms, $\tilde{f}$ can be defined on each $B\in\mathfrak{B}$ as the sum of the valued assumed on the atoms $A_1^{\varepsilon_1}\cap\ldots\cap A_n^{\varepsilon_n}\subset B$. In this way we obtain a unique extension of $\tilde{f}$ on $\mathfrak{B}$ which is additive on disjoint elements and normalized by construction (see Eq. (\ref{eq:sum2})). It follows that such an extension is a measure $\Longleftrightarrow$ it is non-negative $\Longleftrightarrow$ $\tilde{f}$ is non-negative on the elements of $\widetilde{X}$. By construction of $\tilde{f}$, see Eq. (\ref{eq:misat}), such a condition is equivalent to the system of inequalities (\ref{eq:hrep}).

To conclude, we just have to show that the function $f$ defined above takes values in $[0,1]$. Since inequalities (\ref{eq:hrep}) amount to the condition $\tilde{f}\geq 0$, by Eq. (\ref{eq:sum}), it follows that $f\geq 0$. This implies, by Eq. (\ref{eq:sum2}), also that $f\leq 1$.\hfill \endproof


We now prove that the above inequalities are tight, i.e., they cannot be represented as a positive weighted sum of other inequalities and define facets of the polytope. First, we need the following


\begin{lemma}\label{lemma:full} $\mathcal{CP}_n$ is full dimensional in $\mathbb{R}^{2^n-1}$ for all $n$.
\end{lemma}


\textbf{Proof} Given $p\in\mathcal{CP}_n$, it can be written, with the notation of Proposition \ref{prop:polit}, as $p=\sum_\varepsilon \lambda(\varepsilon)u_\varepsilon$.
Now, consider the vectors $p',u'_\varepsilon,\overline{\lambda}\in \mathbb{R}^{2^n}$ defined as $p'=(p,1)$, $u'_\varepsilon=(u_\varepsilon,1)$ and $\overline{\lambda}$ as the vector given by the $2^n$ values $\lambda(\varepsilon)$ for $\varepsilon \in \{0,1\}^n$.
Notice that, by Lemma \ref{lemma:normmeas}, the vector $\overline{\lambda}$ is given by the values assumed by $\tilde{f}$

Given $p\in\mathcal{CP}_n$, by definition of $\mathcal{CP}_n$ and Eq. (\ref{eq:sum2}), there exists a $2^n\times 2^n$ matrix $M$, which columns are given by the vectors $u'_\varepsilon$, such that $p'=M\overline{\lambda}$. On the other hand, by definition of $\tilde{f}$, each component of $\overline{\lambda}$ can be written as a linear combination of the components of $p'$, i.e., there exists a matrix $N$ such that $\overline{\lambda}=N p'$. Since the set of admissible values for $\overline{\lambda}$, see proof of Theorem \ref{teo:hr}, contains a basis for $\mathbb{R}^{2^n}$, e.g., the canonical basis, it follows that $N=M^{-1}$.

As a consequence, the columns of $M$, i.e., the vectors $u'_\varepsilon$, are linearly independent. This implies that the dimension of the subspace spanned by $u_\varepsilon$ is $2^{n}-1$.\hfill \endproof


In order to show that the inequalities (\ref{eq:hrep}) are tight, it is sufficient to notice that $\mathcal{CP}_n$ has exactly $2^{n}$ facets. It is sufficient to count the number of affine hyperplanes, i.e., affine subspaces of dimension $2^n-2$, generated by the vertices $u_\varepsilon$. By the above lemma, the vectors $u_\varepsilon$ are affinely independent, therefore each subset of $2^n-1$ vectors defines an affine hyperplane. Therefore, the number of facets of $\mathcal{CP}_n$ is ${2^n \choose 2^n-1}=2^n$.


\end{document}